\documentclass[twocolumn,twoside,slac_two]{revtex4}
\usepackage{graphicx}
\usepackage{fancyhdr}
\usepackage{amsmath}
\begin{document}

%\title{\boldmath $D^0\bar D^0$ Quantum Correlations, Mixing, and Strong Phases}
\title{\boldmath Measurement of the Strong Phase in $D^0\to K^+\pi^-$ Using
Quantum Correlations}


\author{W. M. Sun (for the CLEO Collaboration)}
\affiliation{Laboratory for Elementary-Particle Physics, Cornell University, Ithaca, NY 14853, USA}

\begin{abstract}
We exploit the quantum coherence between pair-produced $D^0$ and
$\bar D^0$ in $\psi(3770)$ decays to study charm mixing and to make a first
measurement of the relative strong phase $\delta$ between $D^0\to K^+\pi^-$ and
$\bar D^0\to K^+\pi^-$.  Using 281 ${\rm pb}^{-1}$ of $e^+e^-$ collision data
collected with the
CLEO-c detector at $E_{\rm cm}=3.77$ GeV, as well as branching fraction
input from other experiments, we make a preliminary determination of
$\cos\delta = 1.03\pm 0.19\pm 0.08$, where the uncertainties are
statistical and systematic, respectively.  By further including
other external mixing parameter measurements, we obtain an alternate
measurement of $\cos\delta = 0.93\pm 0.32\pm 0.04$, where the systematic
uncertainty from assuming $x\sin\delta=0$ has not been included.
\end{abstract}

\maketitle

\thispagestyle{fancy}


\section{Introduction}
Recent measurements of $D^0$-$\bar D^0$ mixing
parameters~\cite{kpiBelle,ycpBelle,kspipiBelle,kpiBABAR}
highlight the need for information on the relative phase between
the Cabibbo favored decay $D^0 \to K^-\pi^+$ and the doubly Cabibbo
suppressed decay $\bar D^0 \to K^-\pi^+$.  Here, we present a measurement that
takes advantage of the correlated production of $D^0$ and $\bar D^0$ mesons 
in $e^+e^-$ collisions.
If there are no accompanying particles, the $D^0\bar D^0$ pair is in a
quantum-coherent $C=-1$ state.  Because the initial state (the virtual photon)
has $J^{PC} = 1^{--}$, there follows a set of selection rules for the decays
of the $D^0$ and
$\bar D^0$~\cite{Kingsley:1975fe,Okun:1975di,Kingsley:1976cd,Goldhaber:1976fp,Bigi:1986dp,Bigi:1986rj,Bigi:1989ah,Xing:1996pn,Gronau:2001nr,Bianco:2003vb,Atwood:2002ak}.  For example,
both $D^0$ and $\bar D^0$
cannot decay to $CP$ eigenstates with the same eigenvalue.  On the other hand,
decays to $CP$ eigenstates of opposite eigenvalue are enhanced by a factor of
two.
More generally, final states that can be reached by both $D^0$ and $\bar D^0$
(such as $K^-\pi^+$) are subject to similar interference effects.  As a result,
the effective $D^0$ branching fractions in this $D^0\bar D^0$ system differ
from those measured in isolated $D^0$ mesons.  Moreover, using time-independent
rate measurements, it becomes possible to probe $D^0$-$\bar D^0$ mixing as
well as the relative strong phases between $D^0$ and $\bar D^0$ decay
amplitudes to any given final state.

In the Standard Model, $D^0$-$\bar D^0$ mixing is suppressed both by the GIM
mechanism and by CKM matrix elements, although sizeable mixing could arise
from new physics~\cite{Bianco:2003vb}.
Charm mixing is conventionally described by two small dimensionless parameters:
\begin{eqnarray}
x &=& 2\frac{M_2 - M_1}{\Gamma_2 + \Gamma_1} \\
y &=& \frac{\Gamma_2 - \Gamma_1}{\Gamma_2 + \Gamma_1},
\end{eqnarray}
where $M_{1,2}$ and $\Gamma_{1,2}$ are the masses and widths, respectively,
of the neutral $D$ meson $CP$ eigenstates, $D_1$ ($CP$-odd) and
$D_2$ ($CP$-even), which are defined as follows:
\begin{eqnarray}
|D_1\rangle \equiv \frac{|D^0\rangle + |\bar D^0\rangle}{\sqrt{2}} \\
|D_2\rangle \equiv \frac{|D^0\rangle - |\bar D^0\rangle}{\sqrt{2}},
\end{eqnarray}
assuming $CP$ conservation.  The mixing probability is then denoted by
$R_{\rm M}\equiv (x^2+y^2)/2$, and the width of the $D^0$ and $\bar D^0$
flavor eigenstates is $\Gamma\equiv (\Gamma_1+\Gamma_2)/2$.

Many previous searches for charm mixing have focused on $D^0$ decay
times.  Direct measurements of $y$ come from comparing lifetimes in
$D^0\to K^+K^-$ and $\pi^+\pi^-$ decay to that in $D^0\to K^-\pi^+$.
An indirect measure of $y$ is provided by the
``wrong-sign'' process $D^0\to K^+\pi^-$, where
interference between the doubly-Cabibbo-suppressed (DCS) amplitude and the
mixing amplitude manifests itself in the apparent $D^0$ lifetime.
These analyses are sensitive to $y'\equiv y\cos\delta - x\sin\delta$, where
$-\delta$ is the phase of the amplitude ratio
$\langle K^+\pi^-|D^0\rangle / \langle K^+\pi^-|\bar D^0\rangle$.  Below, we
also denote the magnitude of this ratio by $r$, which is measured to
be approximately 0.06.  Because $\delta$ has not previously been measured,
the separate determinations of $y$ and $y'$ above have not been directly
comparable.

In this note, we present an implementation of the method described in
Ref.~\cite{Asner:2005wf} for measuring $y$ and $\cos\delta$ using quantum
correlations at the $\psi(3770)$ resonance.
Our experimental technique is an extension of the double tagging method
previously used to determine absolute hadronic $D$-meson branching fractions
at CLEO-c~\cite{dhad56}.
This method combines yields of fully-reconstructed single tags (ST),
which are individually reconstructed $D^0$ or $\bar D^0$ candidates,
with yields of double tags (DT), which are events where both $D^0$ and
$\bar D^0$ are reconstructed, to
give absolute branching fractions without needing to know the luminosity or
$D^0\bar D^0$ production cross section.  Given a set of input yields,
efficiencies, and background estimates, a least-squares
fitter~\cite{Sun:2005ip}
extracts the number of $D^0\bar D^0$ pairs produced (${\cal N}$) and the
branching fractions (${\cal B}$) of the reconstructed $D^0$ final states,
while accounting for all statistical and systematic uncertainties and their
correlations.
We employ a modified version of this fitter that also determines
$y$, $x^2$, $r^2$, and $r\cos\delta$
using the following categories of
reconstructed final states:  $D\to K^\mp\pi^\pm$,
$CP$-even ($S_+$) and $CP$-odd ($S_-$) eigenstates, and
semileptonic decays ($e^\pm$).  For optimal precision on $\delta$, we also
incorporate measurements of branching fractions and mixing parameters
from other CLEO-c analyses or from external sources.
$CP$ violation in
$D$ and $K$ decays are negligible second order effects that we ignore.
%We neglect $CP$ violation in $D$ decays, which
%would entail a slight correction to the mixing signal.

\section{Formalism}\label{sec:formalism}

To first order in $x$ and $y$, the $C$-odd width $\Gamma_{D^0\bar D^0}(i,j)$
for $D^0\bar D^0$ decay
to final state $i/j$ follows from the anti-symmetric amplitude ${\cal M}_{ij}$:
\begin{eqnarray}\label{eq:rates}
\nonumber
\Gamma_{D^0\bar D^0}(i,j) &\propto& {\cal M}^2_{ij} =
        \left| A_i \bar A_j - \bar A_i A_j \right|^2 \\
	&=& \left|\langle i|D_2\rangle\langle j|D_1\rangle -
        \langle i|D_1\rangle\langle j|D_2\rangle \right|^2,
\end{eqnarray}
where $A_i\equiv\langle i|D^0\rangle$,
$\bar A_i\equiv\langle i|\bar D^0\rangle$.
The total width, $\Gamma_{D^0\bar D^0}$, is the same as for uncorrelated decay,
as are ST rates.
However, unlike the case of uncorrelated $D^0\bar D^0$, we can consider the
$C$-odd $D^0\bar D^0$ system as a $D_1D_2$ pair.
If only flavored final states are considered, as in Ref.~\cite{dhad56},
then the effects of quantum correlations are negligible.  In this
analysis, we also include $CP$
eigenstates, which brings additional sensitivity to $y$ and $\delta$, as
demonstrated below.

Quantum-correlated semileptonic rates probe $y$ because the decay
width does not depend on the $CP$ eigenvalue of the parent $D$ meson,
as this weak decay is only sensitive to flavor content.  However, the total
width of the parent meson does depend on its $CP$ eigenvalue:
$\Gamma_{1,2}=\Gamma(1\mp y)$, so the semileptonic branching fraction
for $D_1$ or $D_2$ is modified by $1\pm y$.  If we reconstruct a semileptonic
decay in the
same event as a $D_2\to S_+$ decay, then the semileptonic $D$ must be a $D_1$.
Therefore, the effective quantum-correlated $D^0\bar D^0$ branching fractions
(${\cal F}^{\rm cor}$) for $CP$-tagged semileptonic final states depend on $y$:
\begin{equation}
{\cal F}^{\rm cor}_{S_\pm/\ell} \approx
	2{\cal B}_{S_\pm}{\cal B}_{\ell}(1\pm y).
\end{equation}
Combined with estimates of ${\cal B}_\ell$ and ${\cal B}_{S_\pm}$
from ST yields, external sources, and flavor-tagged semileptonic yields,
this equation allows $y$ to be determined.

Similarly, if we reconstruct a $D\to K^-\pi^+$ decay in the same event
as a $D_2\to S_+$, then we know the $K^-\pi^+$ was produced from a $D_1$.
The effective branching fraction for this DT process is therefore
\begin{eqnarray}
\nonumber
{\cal F}^{\rm cor}_{S_+/K\pi} &=&
|\langle S_+|D_2\rangle \langle K^-\pi^+|D_1\rangle|^2 \\
\nonumber
&=& A_{S_+}^2 |A_{K^-\pi^+} + \bar A_{K^-\pi^+}|^2 \\
\nonumber
&=& A_{S_+}^2 A_{K^-\pi^+}^2 |1+re^{-i\delta}|^2 \\
&\approx& {\cal B}_{S_+}{\cal B}_{K\pi}(1+R_{\rm WS}+2r\cos\delta+y),
\end{eqnarray}
where $R_{\rm WS} \equiv \Gamma(\bar D^0\to K^-\pi^+)/\Gamma(D^0\to K^-\pi^+) = r^2 + ry' + R_{\rm M}$, and we have used 
${\cal B}_{S_\pm}\propto A_{S_\pm}^2(1\mp y)$ and
${\cal B}_{K\pi}\propto A_{K^-\pi^+}^2(1+ry\cos\delta+rx\sin\delta)$.
In an analogous fashion, we find
${\cal F}^{\rm cor}_{S_-/K^-\pi^+}\approx{\cal B}_{S_-}{\cal B}_{K\pi}(1+R_{\rm WS}-2r\cos\delta-y)$.
When combined with knowledge of ${\cal B}_{S_+}$, $y$, and
$r$, the asymmetry between these two DT yields gives $\cos\delta$.
In the absence of quantum correlations, the effective branching fractions above
would be ${\cal B}_{S_\pm}{\cal B}_{K\pi}(1+R_{\rm WS})$.

More concretely, we evaluate Eq.~\ref{eq:rates} with
the above definitions of $r$
and $\delta$ to produce the expressions in Table~\ref{tab:rates}.
In doing so, we use the fact that inclusive ST rates
are given by the incoherent branching fractions since each event contains
one $D^0$ and one $\bar D^0$.
%integrating over the
%final states of the other side effectively removes the object with which
%the ST is correlated, so the ST behaves as if it were uncorrelated.
Comparison of ${\cal F}^{\rm cor}$ with the uncorrelated effective
branching fractions, ${\cal F}^{\rm unc}$, also given in
Table~\ref{tab:rates}, allows us to extract $r^2$, $r\cos\delta$, $y$, and
$x^2$.
Information on ${\cal B}_i$ is obtained from ST yields at the $\psi(3770)$
and from external measurements using incoherently-produced $D^0$ mesons.
These two estimates of ${\cal B}_i$ are averaged by the fitter to obtain
${\cal F}^{\rm unc}$.

\begin{table}[htb]
\caption{Correlated and uncorrelated effective $D^0\bar D^0$ branching
fractions, ${\cal F}^{\rm cor}$ and
${\cal F}^{\rm unc}$, to leading order in $x$,
$y$ and $r^2$, divided by ${\cal B}_i$
for ST modes $i$ (first section) and
${\cal B}_i{\cal B}_j$ for DT modes $i/j$ (second section).
Charge conjugate modes are implied.}
\label{tab:rates}
\begin{tabular}{ccc}
\hline\hline
Mode & $C$-odd & Uncorr. \\
\hline
$K^-\pi^+$ &
        $1+R_{\rm WS}$ &
        $1+R_{\rm WS}$ \\
$S_+$ & $2$ & $2$ \\
$S_-$ & $2$ & $2$ \\
\hline
$K^-\pi^+$/$K^-\pi^+$ &
        $R_{\rm M}$ &
        $R_{\rm WS}$ \\
$K^-\pi^+$/$K^+\pi^-$ &
        $(1+R_{\rm WS})^2-4r\cos\delta(r\cos\delta+y)$ &
        $1+R_{\rm WS}^2$ \\
$K^-\pi^+$/$S_+$ &
        $1+ R_{\rm WS}+2r\cos\delta+y$ &
        $1+R_{\rm WS}$ \\
$K^-\pi^+$/$S_-$ &
        $1+R_{\rm WS}- 2r\cos\delta-y$ &
        $1+R_{\rm WS}$ \\
$K^-\pi^+$/$e^-$ &
        $1-ry\cos\delta-rx\sin\delta$ &
        $1$ \\
$S_+$/$S_+$ & 0 & $1$ \\
$S_-$/$S_-$ & 0 & $1$ \\
$S_+$/$S_-$ &
        $4$ &
        $2$ \\
$S_+$/$e^-$ &
        $1+y$ &
        $1$ \\
$S_-$/$e^-$ &
        $1-y$ &
        $1$\\
\hline\hline
\end{tabular}
\end{table}

\section{Fit Inputs}

We analyze 281 ${\rm pb}^{-1}$ of $e^+e^-$
collision data produced by the Cornell Electron Storage Ring (CESR) at
$E_{\rm cm}=3.77$ GeV and collected with the CLEO\nobreakdash-c detector,
which is described in detail elsewhere~\cite{cleodetector}.
We reconstruct the $D^0$ and $\bar D^0$ final states listed in
Table~\ref{tab:finalStates}, with $\pi^0/\eta\to\gamma\gamma$,
$\omega\to\pi^+\pi^-\pi^0$, and $K^0_S\to\pi^+\pi^-$.
Signal and background efficiencies, as well as crossfeed probabilities among
signal modes, are determined from simulated events that are processed in a
fashion similar to data.

\begin{table}[htb]
\caption{$D$ final states reconstructed in this analysis.}
\label{tab:finalStates}
\begin{tabular}{cc}
\hline\hline
Type & Final States \\
\hline
Flavored & $K^-\pi^+$, $K^+\pi^-$ \\
$S_+$ & $K^+K^-$, $\pi^+\pi^-$, $K^0_S\pi^0\pi^0$, $K^0_L\pi^0$ \\
$S_-$ & $K^0_S\pi^0$, $K^0_S\eta$, $K^0_S\omega$ \\
$e^\pm$ & Inclusive $Xe^+\nu$, $Xe^-\bar\nu$ \\
\hline\hline
\end{tabular}
\end{table}

Hadronic final states without $K^0_L$ mesons are fully reconstructed
via two kinematic variables: the beam-constrained candidate mass,
$Mc^2 \equiv\sqrt{E_0^2 - {\mathbf p}_D^2c^2}$, where ${\mathbf p}_D$ is the
$D^0$ candidate momentum and $E_0$ is the beam energy, and
$\Delta E\equiv E_D - E_0$, where
$E_D$ is the sum of the $D^0$ candidate daughter energies.
We extract ST and DT yields from $M$ distributions using
unbinned maximum likelihood fits (ST) or by counting candidates in signal
and sideband regions (DT).

\begin{table}[tb]
\caption{ST and DT yields, efficiencies, and their statistical
uncertainties. For DT yields, we sum groups of modes and provide an
average efficiency for each group;
the number of modes in each group is given in parentheses.
Modes with asterisks are not included in the standard and extended fits.}
\label{tab-STDTYieldsAndEffs}
\begin{tabular}{lcc}
\hline\hline
Mode &  ~~Yield~~ & ~~Efficiency (\%) \\ \hline
$K^-\pi^+$ &
	$25400\pm 200$ & $64.70\pm 0.04$ \\
$K^+\pi^-$ &
	$25800\pm 200$ & $65.62\pm 0.04$ \\
$K^+K^-$ &
	$4740\pm 70$ & $57.25\pm 0.09$ \\
$\pi^+\pi^-$ &
	$2100\pm 60$ & $72.92\pm 0.13$ \\
$K^0_S\pi^0\pi^0$ &
	$2440\pm 70$ & $12.50\pm 0.06$ \\
$K^0_S\pi^0$ &
	$7520\pm 90$ & $29.73\pm 0.05$ \\
$K^0_S\eta$ &
	$1050\pm 40$ & $10.34\pm 0.06$ \\
$K^0_S\omega$ &
	$3240\pm 60$ & $12.48\pm 0.04$ \\
\hline
$K^\mp\pi^\pm$/$K^\mp\pi^\pm$ (2) & $4\pm 2$ & $40.2\pm 2.4$ \\
$K^-\pi^+$/$K^+\pi^-$ (1) & $600\pm 25$ & $41.1\pm 0.2$ \\
$K^\mp\pi^\pm$/$S_+$ (8) & $605\pm 25$ & $26.1\pm 0.1$ \\
$K^\mp\pi^\pm$/$S_-$ (6) & $243\pm 16$ & $12.3\pm 0.1$ \\
$K^\mp\pi^\pm$/$e^\mp$ (2) & $2346\pm 65$ & $45.6\pm 0.1$ \\
$S_+$/$S_+$ (9*) & $10\pm 6$ & $12.5\pm 0.6$ \\
$S_-$/$S_-$ (6*) & $2\pm 2$ & $3.9\pm 0.2$ \\
$S_+$/$S_-$ (12) & $242\pm 16$ & $7.7\pm 0.1$\\
$S_+$/$e^\mp$ (6) & $406\pm 44$ & $22.2\pm 0.1$ \\
$S_-$/$e^\mp$ (6) & $538\pm 40$ & $13.8\pm 0.1$ \\
\hline\hline
\end{tabular}
\end{table}

Because most $K^0_L$ mesons and neutrinos produced at CLEO-c are not
detected,
we only reconstruct modes with these particles in DTs, by demanding that the
other $D$ in the event be fully reconstructed.
%modes with these particles are only reconstructed in double tags, paired with a
%fully-reconstructed hadronic $D$ decay.  
Ref.~\cite{ksklpi} describes the missing mass technique used to identify
$K^0_L\pi^0$ candidates.
For semileptonic decays, we use inclusive, partial reconstruction to maximize
efficiency, demanding that only the electron be identified with a
multivariate discriminant~\cite{eid} that combines measurements from the 
tracking
chambers, the electromagnetic calorimeter, and the ring-imaging
\v{C}erenkov counter.

Table~\ref{tab-STDTYieldsAndEffs} gives yields and efficiencies for
8 ST modes and 58 DT modes, where the DT modes have been grouped into
categories.  Fifteen of
the DT modes are forbidden by $CP$ conservation and are not included in
the standard fit.
In general, crossfeed among signal modes and backgrounds from other $D$ decays
are smaller than 1\%.  Modes with $K^0_S\pi^0\pi^0$ have approximately
3\% background, and yields for $K^\mp\pi^\pm$/$K^\mp\pi^\pm$ and
$S_\pm/S_\pm$ are consistent with being entirely from background.

External inputs to the fit include measurements of $R_{\rm M}$, $R_{\rm WS}$,
${\cal B}_{K^-\pi^+}$, and ${\cal B}_{S_\pm}$, as well as an
independent ${\cal B}_{K^0_L\pi^0}$ from CLEO-c, as shown in
Table~\ref{tab:externalMeas1}.  The external $R_{\rm WS}$ is required to
constrain $r^2$, and thus, to determine $\cos\delta$
from $r\cos\delta$.
We also use the external mixing
parameter measurements shown in Table~\ref{tab:externalMeas2}.
The fit incorporates the full covariance matrix for these inputs,
accounting for statistical overlap with the yields in this analysis.
Covariance matrices for the fits in Ref.~\cite{wskpi} have been provided by
the CLEO, Belle, and BABAR collaborations.

\begin{table}[htb]
\begin{center}
\caption{Averages of external measurements used in the standard fit.
Charge-averaged $D^0$ branching fractions are denoted by final state.}
\label{tab:externalMeas1}
\begin{tabular}{lc}
\hline\hline
Parameter & Average \\
\hline
$R_{\rm WS}$ & $0.00409\pm 0.00022$~\cite{rws}  \\
$R_{\rm M}$ & $0.00017\pm 0.00039$~\cite{rm} \\
$K^-\pi^+$   & $0.0381\pm 0.0009$~\cite{pdg04} \\
$K^-K^+/K^-\pi^+$     & $0.1010\pm 0.0016$~\cite{pdg06} \\
$\pi^-\pi^+/K^-\pi^+$ & $0.0359\pm 0.0005$~\cite{pdg06} \\
$K^0_L\pi^0$  & $0.0097\pm 0.0003$~\cite{ksklpi} \\ %Incl. pi0 corr + error
$K^0_S\pi^0$ & $0.0115\pm 0.0012$~\cite{pdg04} \\
$K^0_S\eta$  & $0.00380\pm 0.00060$~\cite{pdg04} \\
$K^0_S\omega$  & $0.0130\pm 0.0030$~\cite{pdg04} \\
\hline\hline
\end{tabular}
\end{center}
\end{table}

\begin{table}[htb]
\begin{center}
\caption{Averages of external measurements used in the standard and
extended fits.}
\label{tab:externalMeas2}
\begin{tabular}{lc}
\hline\hline
Parameter & Average \\
\hline
$y$    & $0.00662\pm 0.00211$~\cite{pdg06,ycpBelle,kspipi} \\
$x$    & $0.00811\pm 0.00334$~\cite{kspipi} \\
$r^2$  & $0.00339\pm 0.00012$~\cite{wskpi} \\
$y'$   & $0.0034\pm 0.0030$~\cite{wskpi} \\
$x'^2$ & $0.00006\pm 0.00018$~\cite{wskpi} \\
\hline\hline
\end{tabular}
\end{center}
\end{table}

Systematic uncertainties associated with efficiencies for reconstructing
tracks, $K^0_S$ decays, $\pi^0$ decays, and for hadron identification
are assigned as described in Ref.~\cite{dhadprd}.
Other sources of efficiency uncertainty include:
$\Delta E$ requirements (0.5--5.5\%),
$\eta$ reconstruction (4.0\%),
electron identification (1.0\%),
modeling of particle multiplicity and detector noise (0.1--1.3\%),
simulation of initial and final state radiation (0.5--1.2\%),
and modeling of resonant substructure in $K^0_S\pi^0\pi^0$ (0.7\%).
We also include additive
uncertainties of 0.0--0.9\% to account for variations of yields with 
fit function.

These systematic uncertainties are included directly in the
covariance matrix given to the fitter, which propagates them to the fit
parameters.
The other fit inputs determined in this analysis are
ST and DT yields and efficiencies, crossfeed probabilities,
background branching fractions and efficiencies, and statistical
uncertainties on all of these measurements.  Quantum correlations between
signal and background modes are accounted for using assumed values of
amplitude ratios and strong phases that are systematically varied and found
to have negligible effect.
We validated our analysis technique in a simulated $C$-odd
$D^0\bar D^0$ sample 15 times the size of our data sample.

\section{Preliminary Fit Results}

Our standard fit excludes
the 15 same-$CP$ DT modes and includes the measurements in
Table~\ref{tab:externalMeas1} but not Table~\ref{tab:externalMeas2}.  
In this fit, there is not enough information to reliably determine
$x\sin\delta$, so we fix it to zero, and the associated systematic
uncertianty is negligible.
We obtain a first measurement of $\cos\delta=1.03\pm 0.19\pm 0.08$,
consistent with being at the boundary of the physical region.
The fit results for $y$, $r^2$, $x^2$, and branching fractions are
consistent with previous measurements.

\begin{figure}[htb]
\centering
\includegraphics*[width=\linewidth]{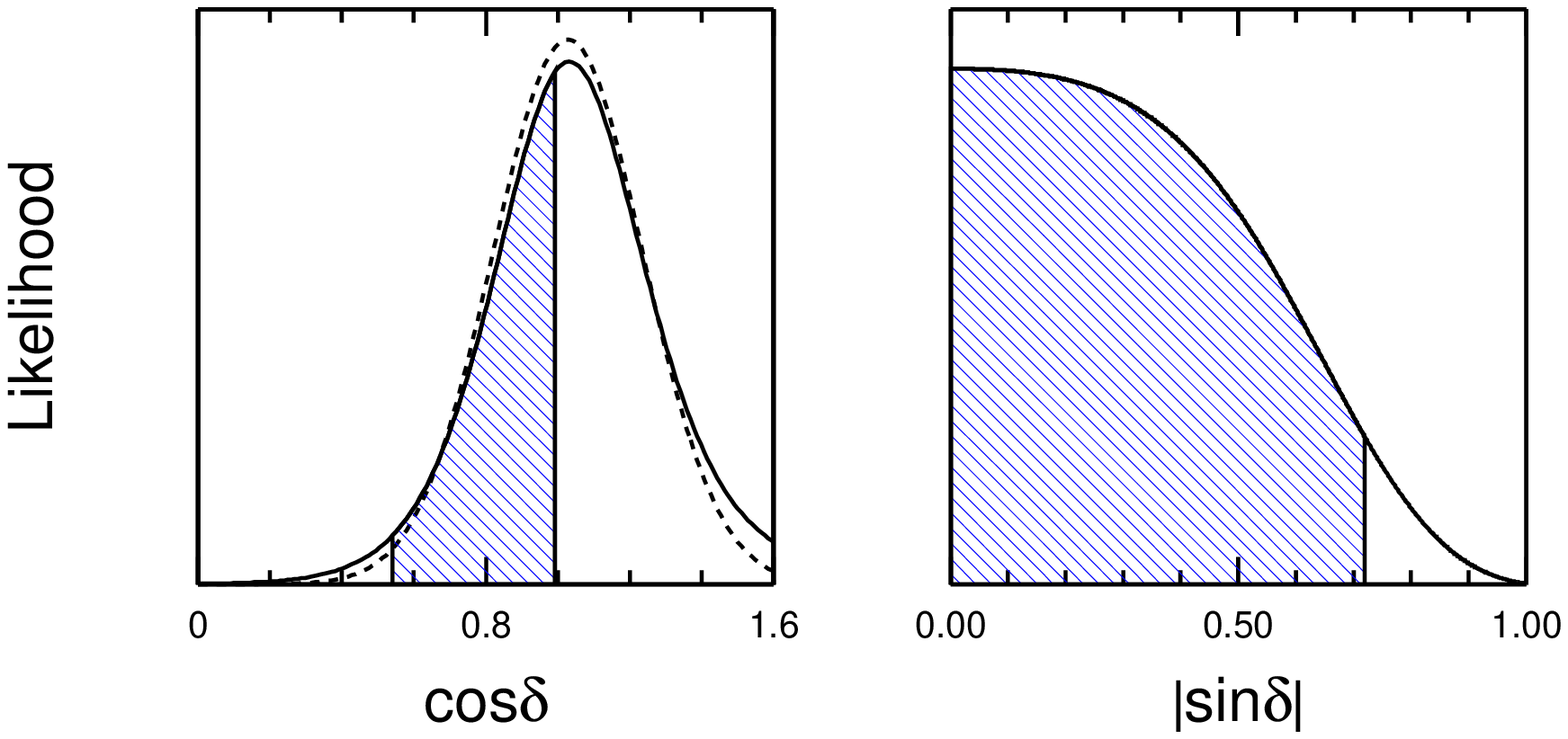}
\includegraphics*[width=\linewidth]{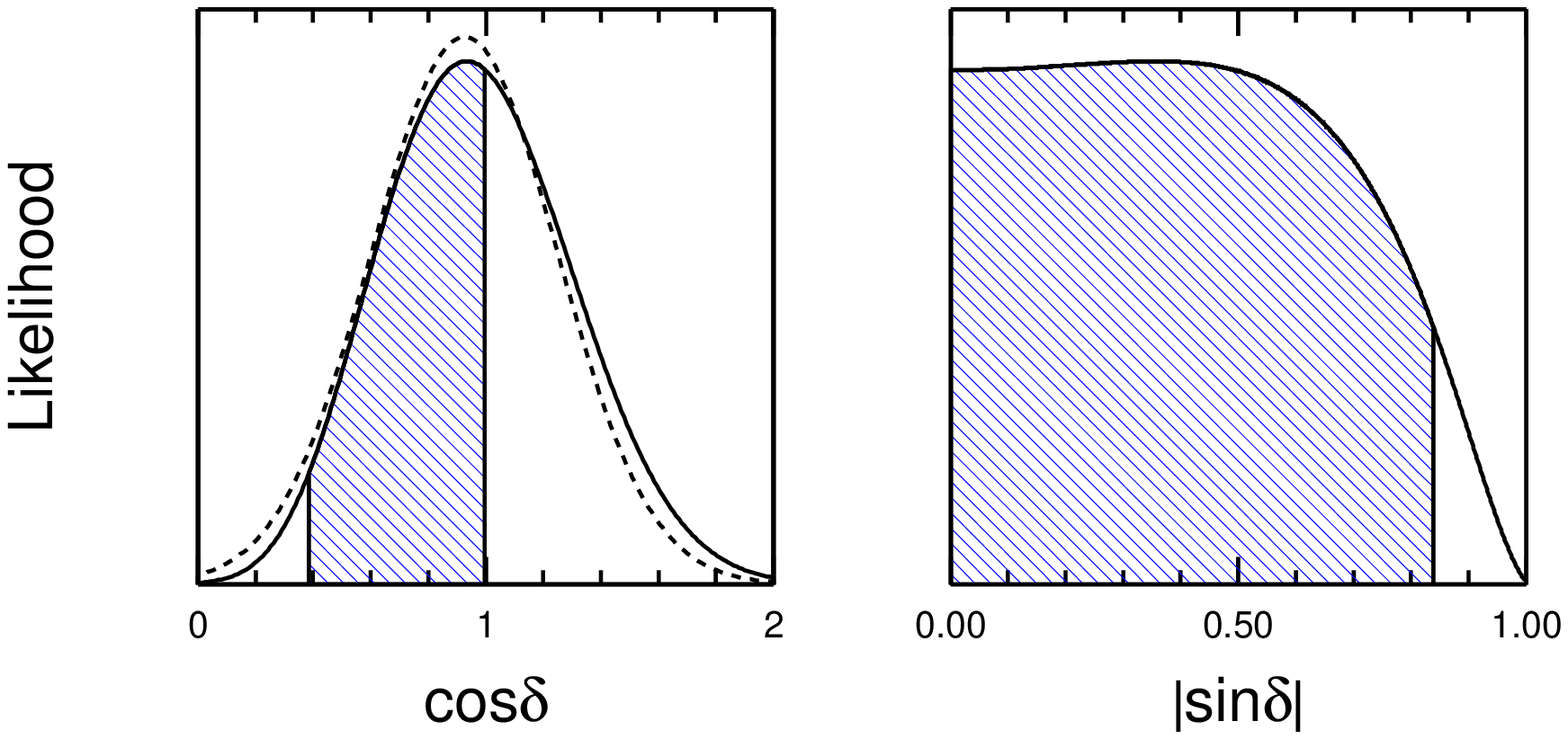}
\caption{Standard (top) and extended (bottom)
fit likelihood functions for $\cos\delta$ and $|\sin\delta|$, including
both statistical and systematic uncertainties, excluding the uncertainty
from assuming $x\sin\delta=0$.  The dashed curve shows the
Gaussian likelihood corresponding to the standard fit result.  The hatched
regions contain 95\% of the areas in the physical regions.}
\label{fig:fitLikelihoods}
\end{figure}

The likelihood curve for $\cos\delta$, computed as
${\cal L}=e^{-(\chi^2-\chi^2_{\rm min})/2}$ and shown in
Figure~\ref{fig:fitLikelihoods},
is slightly non-Gaussian.  For values of $|\cos\delta| < 1$, we also
show ${\cal L}$ as a function of $|\sin\delta|$.  We integrate these
curves within the physical region to obtain 95\% confidence level limits
of $\cos\delta > 0.54$ and $|\sin\delta| < 0.72$.

We also perform an extended fit that includes the previous measurements of
$y$ and $y'$ in Table~\ref{tab:externalMeas2}, in addition to all the
inputs to the standard fit above.  In this fit, we find
$\cos\delta=0.93\pm 0.32\pm 0.04$.  The systematic uncertainty does not
include the contribution from assuming $x\sin\delta=0$, which is still
under study.  From the corresponding likelihood functions shown in
Figure~\ref{fig:fitLikelihoods}, we determine 95\% confidence level limits
of $\cos\delta > 0.38$ and $|\sin\delta| < 0.84$.

The $\cos\delta$ uncertainty in the extended fit is larger than in the
standard fit because of
a non-linear effect.  Most of the information on $r^2$ (and therefore on $r$)
is provided by $R_{\rm WS}$.  Because $R_{\rm WS}$ also depends on
$y\cdot r\cos\delta$, the sign of the
correlation between $r^2$ and $r\cos\delta$ is given by the sign of $y$.
In the standard fit, $y$ attains a more negative central value than in
the extended fit, where $y$ is constrained to the precise external
measurements.  Hence, the uncertainty on $\cos\delta$ becomes inflated
in the extended fit.

By observing the change in $1/\sigma_{\cos\delta}^2$ as
each fit input is removed, we identify the major contributors of
information about $\cos\delta$ to be the $K\pi/S_\pm$ DT yields and the
ST yields.
We also find that no single input or group of inputs exerts a pull larger
than three standard deviations on $\cos\delta$ or $y$.  Moreover, removing
all external inputs gives branching
fractions consistent with those in Table~\ref{tab:externalMeas1}.

We also allow for a $C$-even $D^0\bar D^0$ admixture in the initial state,
which is expected to be ${\cal O}(10^{-8})$~\cite{petrov}, by including
the 15
$S_\pm/S_\pm$ DT yields in the fit. These modes limit the $C$-even component,
which can modify the other yields as described in
Ref.~\cite{Asner:2005wf}.  In both the standard and extended fits,
we find a $C$-even fraction consistent with zero
with an uncertainty of 2.4\%, and neither the fitted $\cos\delta$ values nor
their uncertainties are shifted noticeably from the results quoted above.

\section{Summary}

Using 281 ${\rm pb}^{-1}$ of $e^+e^-$ collisions produced at the
$\psi(3770)$, we make a preliminary first determination of the strong 
phase $\delta$, with $\cos\delta = 1.03\pm 0.19\pm 0.08$.
By further including external
mixing parameter measurements in our analysis, we obtain an alternate
measurement of $\cos\delta = 0.93\pm 0.32\pm 0.04$, where the systematic
uncertainty from assuming $x\sin\delta=0$ has not been included.
Knowledge of $\delta$ allows independent measurements of $y$ and $y'$ to
be combined, thereby improving our overall knowledge of charm mixing
parameters.

\begin{acknowledgments}
We gratefully acknowledge the effort of the CESR staff in providing
us with excellent luminosity and running conditions.  This work was
supported by the National Science Foundation and the U.S. Department
of Energy.
\end{acknowledgments}

\bigskip % extra skip inserted
%\begin{thebibliography}{9}   % Use for  1-9  references

\end{document}